\documentclass[aps,prl,twocolumn,groupedaddress, showpacs]{revtex4}

 \usepackage{graphicx}
 \usepackage{amssymb}
 \usepackage{epstopdf}
\begin{document}

\preprint{LBNL-51xxx}
\title{Demonstration of a Cold Atom Fountain Electron Electric Dipole Moment Experiment} 


\author{Jason M. Amini}
 \email{JAMaddi@lbl.gov}
 \altaffiliation[Also at ]{Physics Department, University California at Berkeley, Berkeley, CA 94720.}
  \author{Charles T. Munger Jr.}
 \email{Charles@SLAC.stanford.edu}
 \altaffiliation[Also at ]{SLAC, 2575 Sand Hill Road, Mailstop 59 Menlo Park, CA 94025}
\author{Harvey Gould}
 \email{HAGould@lbl.gov}
\affiliation{Mail stop 71-259, Lawrence Berkeley National Laboratory,  Berkeley, CA 94720}


\date{\today}

\begin{abstract}
 A cesium fountain electron electric dipole moment (e-EDM) experiment using electric-field quantization is demonstrated. With magnetic fields reduced to $B \le$ 200 pT, the electric field $E$ lifts the degeneracy between different $|m_F|$ hyperfine sublevels and, along with the low velocity and fountain geometry, suppresses systematics from the motional magnetic field. Transitions are induced and the atoms polarized and analyzed in field-free regions ($B \le$ 200 pT, $E = 0$). The feasibility of reaching a sensitivity to an e-EDM of $2 \times 10^{-50}$ C-m ($1.3 \times 10^{-29}$ e-cm) in a cesium fountain experiment is discussed.
 \end{abstract}


\maketitle
The observation of neutrino flavor symmetry violation and its association with neutrino mass shows that there is new physics beyond the Standard Model \cite{mohapatra05}. Observing an electron electric dipole moment (e-EDM) would uncover additional new physics, while merely improving the present e-EDM limit would place constraints on current models of neutrino physics \cite{mohapatra05} and other extensions of the Standard Model. Furthermore, it appears feasible to reach an e-EDM sensitivity of $2 \times10^{-50}$ C-m ($1.3 \times 10^{-29}$ e-cm), about two orders of magnitude below recent experiments \cite{regan02, hudson02, abdullah90, murthy89}, using a fountain of cold atoms. 

Potentially observable e-EDMs \cite{bern91, bern91erra} are predicted  by extensions to the Standard Model such as Supersymmetry \cite{abel01}, Multi Higgs models, Left-Right Symmetric models, Technicolor, and others.  Split Supersymmetry \cite{arkani-hamed05, chang05, giudice06} predicts an e-EDM (and a neutron EDM) within a few orders of magnitude of the present experimental limit. And if the matter - antimatter asymmetry in the universe arises from the same mechanism that produces neutrino mass, then e-EDM and (charged) lepton flavor violation experiments may uncover it \cite{mohapatra05}.

Laboratory e-EDM experiments search for a difference in interaction energy of an electron aligned and anti-aligned with an external electric field. High atomic number paramagnetic atoms and molecules provide test systems of zero net charge and may enhance the sensitivity to an e-EDM. The calculated enhancement factor $R$ for the cesium ground state is $114\pm15$ \cite{sandars66, johnson86}. Because the interpretation of the EDM measurement does not depend on subtracting out Standard Model effects, the error in the enhancement factor does not need to be small.

A cesium e-EDM experiment detects the EDM as a shift in the energy between different $m_F$ hyperfine sublevels that is linear in the applied electric field, $E$. To avoid a false positive, symmetry conserving effects that produce linear shifts must be suppressed. The predominant systematic is the motional magnetic field (S.I. units):
\begin{equation}\label{mot}
 \mathbf{B}_{\rm mot} = \mathbf{v \times  E}/c^2
\end{equation}
seen by a neutral atom moving with velocity $v$ through an electric field $E$, where $c$ is the speed of light.
%
\begin{figure}
\includegraphics [scale =0.25] {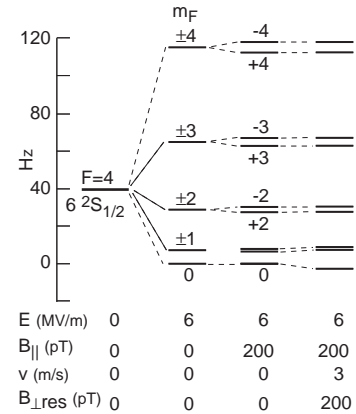}
\caption{\label{energy_levels}
Electric field quantized energy levels of the cesium $6^2S_{1/2}, F = 4$, for small residual and motional magnetic fields (from Eq. \ref{shift}). Each column of numbers represents different experimental conditions.  The conditions in the demonstration experiment are represented by the rightmost column and include a motional magnetic field of 200 pT. }
\end{figure}
%
When a static magnetic field $B_{\rm 0}$ such as may be used to lift the degeneracy between $m_F$ levels, is also present, misalignment between $E$ and $B_{\rm 0}$ causes a component of $B_{\rm mot}$ to lie along $B_{\rm 0}$. The total magnetic field then changes linearly with $E$ and, through the atom's magnetic moment, mimics an EDM \cite{sandars64}.

A fountain EDM experiment can use two effective methods to suppress motional magnetic field effects: atom-by-atom cancellation of the beam velocity by the rise and fall of the atoms under gravity, and electric -ield quantization \cite{player70}, where no static magnetic field is needed because the electric field lifts the degeneracy of sublevels with different $|m_F|$ (Fig. \ref{energy_levels}). 

In erlectric-field quantization, energy shifts due to the motional magnetic field are absent to lowest order. The energy shift $W(m_F)$ of an $F = 4$, $m_F \ne 1$ sublevel in strong electric and weak residual magnetic fields and with the quantization axis defined by the electric field, is given by:
\begin{eqnarray}\label{shift}
\frac{W(m_F)}{h} & = &\epsilon E^2 m_F^2 + g \mu B_{||} m_F \nonumber \\
& &  + K_1 \frac{(g\mu)^2 B_{\perp}^2}{\epsilon E^2} -K_2 \frac{(g\mu)^3 B_{\perp}^2 B_{||}}{(\epsilon E^2)^2}\\
& & -\frac{d_e R m_F E}{4h} + \mbox{higher order terms,}\nonumber
\end{eqnarray} 
where 
$\epsilon = -3 \alpha_T/56$, and
$\alpha_T = -3.5 \times 10^{-12}$ HzV$^{-2}$m$^2$ is the tensor polarizability\cite{gould69, ospelkaus03}, $g\mu \approx 3.5 \times 10^9$ Hz/T, $B_{||}$ is the magnetic field parallel to $E$, and $B_{\perp}$ is the magnetic field perpendicular to $v$ and to $E$, $d_e$ is the e-EDM, $R$ is the enhancement factor, $h$ is Planck's constant, and
\begin{eqnarray}\label{coeff}
K_1(m_F) & = & \frac{m_F^2 +20}{2(4m_F^2 -1)} \\
K_2(m_F)& = & \frac{81 m_F}{2(4m_F^2 -1)^2}. \nonumber
\end{eqnarray}

$B_{\perp}$ includes both $B_{\rm mot}$ and any static residual field $B_{\perp \rm res}$. The leading motional systematic error $W_{\rm sys}(m_F)$ is then from Eq.'s \ref{shift}, \ref{coeff},  
\begin{eqnarray}\label{sys}
\frac{W_{\rm sys}(m_F)}{h} = -2K_2(m_F) \frac{(g\mu)^3 B_{\perp \rm res} B_{\rm mot} B_{||}}{(\epsilon E^2)^2}
\end{eqnarray}
where $B_{\rm mot}$ is given by Eq. \ref{mot} and where $B_{\perp \rm res}$ is taken to be parallel to $B_{\rm mot}$. This term is odd in $E$ (through $B_{\rm mot}$) and odd in $m_F$ (through $K_2$). However, it can be made very small when $E$ and $m_F$ are large and $v$, $B_{\perp \rm res}$, and $B_{||}$ are small.

To test the feasibility of using electric-field quantization in a fountain e-EDM experiment with state preparation, analysis and atom transport in field-free regions,  an existing cesium fountain apparatus \cite{amini03} was modified. Magnetic magnetic shielding with demagnetizing coils were added and inside these were constructed three sets of orthogonal magnetic field coils for nulling residual magnetic fields and for inducing transitions between $m_F$ states.

Even without electric-field quantization, the rise and fall of atoms in a fountain results in an atom-by-atom velocity cancellation that suppresses motional magnetic field effects to where they would be difficult to observe in our low-flux fountain. To magnify motional magnetic field effects and to test electric-field quantization more rigorously, the average beam velocity was set to about 3 m/s by increasing the launch velocity so that the upward traveling atoms did not turn around inside the electric field plates, but instead exited and were analyzed and  detected above the electric field plates.

After launching from the fountain's magneto-optic trap the packet of cesium atoms entered a magnetically shielded and nulled region where the states were prepared, transitions were induced by pulsed fields, the electric field was applied, and the final states analyzed and detected. The atoms were prepared in either the $m_F = + 4$ or -4 state by optical pumping to the $6^2P_{3/2}$,  $F = 4$ state with 852 nm circularly polarized light, and later analyzed to separate those that made a transition out of the $m_F = 4$ ($-4$) sublevel from those that remained. State preparation and analysis were done in the regions free of electric and magnetic fields (B $\le$ 200 pT in each orthogonal direction).

The state analysis used linearly polarized light to excite all but the $m_F = \pm4$ state to the $6^2P_{3/2}$, $F = 3$ level which subsequently decays 75\% of the time to the ground state $F = 3$ hyperfine level. The remaining $25 \%$ of the time an atom returns to the ground $F = 4$ level to be excited again.  After many cycles, the population in the $m_F = \pm 4$ states is the sum of their original populations, plus about 20\% of the population originally in $m_F = \pm 3$ states, plus a smaller percentage of atoms originally in $m_F = \pm 2$ states. 

The atoms left in the $m_F = \pm 4$ states were then counted by exciting the cycling transition $6^2S_{1/2}, F =4 \rightarrow 6^2P_{3/2}, F = 5$ and collecting the fluorescence radiation into a photomultiplier. 
The signal intensity was normalized to the number of atoms present by pumping atoms in the $F = 3$ level back into the $F = 4$ level and again detecting fluorescence from the cycling transition. The low-cost laser, optics, and detection systems had only a single diode laser plus a diode repump laser for trapping, launching, state selection, analysis, and detection. About 1000 atoms per launch were detected.

The residual magnetic fields were mapped in 3 directions at points along the beam using the cesium atoms as a magnetometer (Eq. \ref{shift}). Holes in the magnetic shields (used for access to windows and high-voltage feedthroughs) caused local magnetic field maxima of about 3 nT. Based upon magnetic field mapping, waveform generators were programmed to deliver time-dependent currents to the nulling coils. This produced a local null around the atom packet that followed the atoms as they traveled. The mixing of the $m_F$ states by un-nulled magnetic fields wasstill noticeable but small.

\begin{figure}
\includegraphics [scale =0.5] {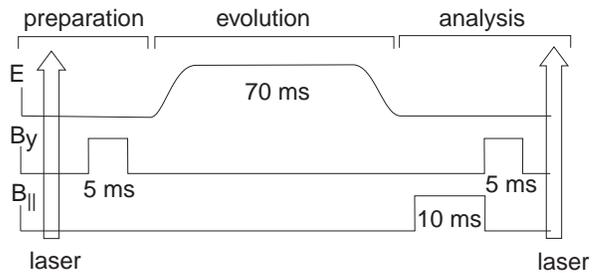}
\caption{\label{timing}
Timing sequence. The mixing ($B_{\rm y}$ and shifting ($B_\parallel$) pulses are applied for a fixed time. The laser beams are at fixed locations.
 }
\end{figure}
\begin{figure}
\includegraphics [scale =0.5] {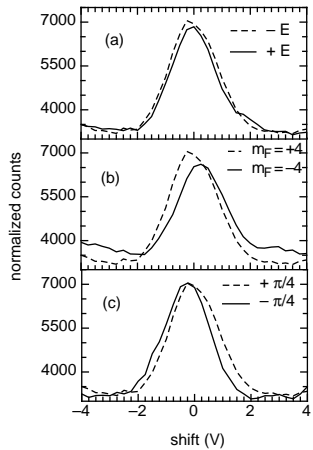}
\caption{\label{resonance}
The $m_F = \pm4$ population as a function of the amplitude of the shifting pulse.  Shown are the effects on the resonance position and shape of: a reversal of the electric field (a);  a change of the initial state $m_F$ between +4 and -4 (b); and a change in polarity of the mixing pulse (c). For reference, the broken line shows a common condition of $E$, $m_F = 4$, and the polarity of the mixing pulse. An $x$-axis value of 2V corresponds to a magnetic field of 100 pT. An e-EDM (or systematic error) of $4 \times 10^{-43}$ C-m ($2.5 \times 10^{-22}$ e-cm) would produce a resonance shift of about 0.1 V.}
\end{figure}

Transitions out of the $m_F = 4$ (-4) state were induced by a sequence of two 5 ms-long square magnetic pulses (``mixing'' pulses) along the direction of motion of the atoms: the first applied before the atoms entered the electric field; and the second after the atoms left the electric field. To tune the transition probability a 10 ms-long magnetic field pulse (``shifting'' pulse), is applied parallel to the electric field after the atoms exit the electric field and before the second mixing pulse (Fig. \ref{timing}). By scanning the amplitude of the shifting pulse, resonance patterns such as the one shown in Fig. \ref{resonance} are generated.

\begin{figure}
\includegraphics [scale =0.5] {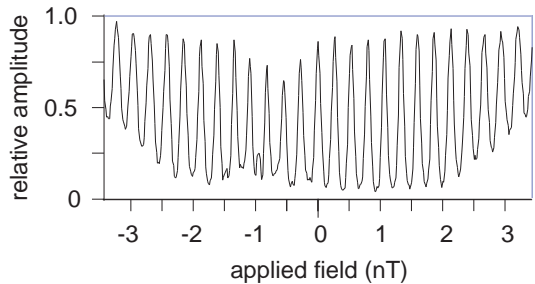}
\caption{\label{scan}
The detected $m_F = \pm4$ population as a function of the amplitude of a constant external magnetic field. The loss of contrast near -0.7 nT  is consistent with a 300 nT remnant magnetic field perpendicular to the electric field. 
 }
\end{figure}

To permit a wider scan for diagnostics purposes, the shifting pulse is replaced by a weak magnetic field applied from state selection through detection. The resulting transition amplitude as a function of magnetic field is shown in Fig. \ref{scan}. The $\approx 5$ Hz width of the resonances is set by the 90 ms transit time of the atoms between mixing pulses. 

In this demonstration experiment, the electric field was set to $\pm 6$ MV/m. Resonance shapes were measured for the two electric field polarities, for the initial states $m_F = 4$ and $m_F = -4$, and for the two polarities of the mixing pulse - a total of eight combinations. Reversing the electric field cancels out the energy shifts that are independent of, or even in, $E$, such as $B_\parallel$ and the tensor Stark shift. Reversing the polarization subtracts terms such as $K_1$ that are even in $m_F$. Reversing the mixing pulses cancels the effects of lineshape distortions to first order.

The line shapes in Fig. \ref{resonance} are not simple sinusoids because the $F = 4$ cesium ground state contains nine sublevels that are coupled by the mixing pulses. An analysis of the lineshapes and their distortions will be presented in a future paper. However because of the distortions, it was important to map out the entire resonance shape as shown in Fig. \ref{resonance}.  To measure energy shifts between the $m_F$ levels, a basic analysis of the data, that does not fully treat distortion effects, was performed.

Eighteen sets of scans were taken, each set with the eight combinations of reversals.  Our centroid analysis obtains an EDM of $-0.7 \pm 2.2 \times 10^{-43}$ C-m ($-0.5 \pm 1.4 \times10^{-22}$ e-cm) where the value in parenthesis is the statistical uncertainty at the $1 \sigma$ level. 

Mapping a set of eight centroids took 40 minutes, leaving the measurement vulnerable to slow drifts in the magnetic field whose effects could otherwise be cancelled by frequent reversals of the electric field. With better magnetic nulling, an apparatus customized for e-EDM experiments, and with the improvements discussed below, future experiments should be able to detect shifts in the resonances by monitoring only a few high-slope points.

We estimate the motional systematic error by considering only the $|m_F|=4$ and 3 states.
For $E = \pm 6$MV/m, $v =3$ m/s, and $B_{||} = B_{\perp \rm res}$ = 200 pT, the systematic (Eq.'s \ref{mot} - \ref{sys}) is $3 \times 10^{-45}$ C-m ($2 \times 10^{-24}$ e-cm) and is not a factor in our demonstration experiment.

In a cesium e-EDM fountain experiment with $E = 13.5$ MV/m, where a rise and subsequent fall of atoms reduces the time-averaged average velocity to $< 3$ mm/s, and where more complete magnetic shielding (and without holes) reduces the residual magnetic fields to 20 pT, the systematic would drop to $1.2 \times 10^{-51}$ C-m ($7 \times 10^{-31}$ e-cm), far below the current experimental limit. 
 
To reach an e-EDM sensitivity of $2 \times10^{-50}$ C-m ($1.3 \times 10^{-29}$ e-cm), 
it is preferable to use a stronger electric field (13.5 MV/m) and a seven-quantum transition $m_F = \pm 4 \leftrightarrow m_F = \mp 3$. This transition is more sensitive to the EDM  (Eq. \ref{shift}) because of the larger change in $m_F$, but requires  more complicated versions of the mixing pulses and state analysis than could be mustered for our demonstration experiment.

Detecting an e-EDM of $2 \times10^{-50}$ C-m in a $13.5$ MV/m electric field requires detecting a frequency shift of  $8 \times10^{-8}$ Hz. With a transit-time limited instrumental linewidth of 1 Hz and an average flux of $2 \times 10^8$ s$^{-1}$ atoms detected, the counting time needed to reach this statistical precision using a seven-quanta transition would require about 225 hours. Measurements to bound systematic errors would require additional time. 

Fluxes of $>1\times10^9$ cesium atoms/s have been launched and cooled to 1.5 $\mu$K or lower \cite{legre98,treutlein01}. To have all or most of these atoms return, it is not sufficient that the atoms be  cold and the electric-field plate gap be large. It is also necessary to focus the atoms to counter the defocusing effect of the electric-field gradient at the entrance of the electric-field plates \cite{maddi99}. 
An electrostatic lens triplet, designed from first principles, has been successfully operated in the same fountain apparatus \cite{kalnins05}. Simulations \cite{kalnins03} show that electrostatic lensing can compensate for the defocusing and produce a near parallel beam between a pair of electric field plates with a 10 mm gap and 13.5 MV/m field.  Nearly 100\% of the atoms return to be detected. Beam flux is reduced by losses in analysis, detection, and by a less than $100$\% seven-photon transition probability.

To reach fields of 13.5 MV/m (or higher) with a 10 mm gap, and especially to reduce magnetic Johnson noise \cite{munger05}, heated glass electrodes can be used. Heated glass electrodes have previously been built and used for polarizability measurements \cite{gould76, marrus69}.  

In conclusion, it appears feasible to reach an e-EDM sensitivity of $2 \times10^{-50}$ C-m using a fountain of cold cesium atoms, seven-quanta transitions, heated glass electrodes, and electrostatic lenses. Motional magnetic field effects would be suppressed by the fountain geometry and by electric field quantization.

\section*{Acknowledgments}
We thank Timothy Dinneen for assistance in constructing the fountain and Douglas McColm for theoretical guidance. We gratefully acknowledge support of a NIST Precision Measurements Grant and support from the NASA Office of Biological and Physical Research. The Lawrence Berkeley National Laboratory is operated for the U.S. DOE under Contract  No. DE-AC02-05CH11231.

\bibliography{EDMbib}
\end{document}